\documentclass[12pt]{iopart_arXiv10062094v6}
\usepackage{iopams}
\usepackage{graphicx}           
\newcommand{\textXXX}{\mathsf}  

\begin{document}  

\title[Newtonian gravity as an entropic force:
       Towards a derivation of $\;G$]
       {Newtonian gravity as an entropic force:\\
       Towards a derivation of $\boldsymbol{G}$}

\author{F.R. Klinkhamer}

\address{Institute for the Physics and Mathematics of the Universe,
              University of Tokyo,
              Kashiwa 277--8583, Japan\\and\\
         Institute for Theoretical Physics, University of Karlsruhe,
             Karlsruhe Institute of Technology, 76128 Karlsruhe,
             Germany \footnote[2]{Permanent address.}
             }  
\ead{frans.klinkhamer@kit.edu}

\begin{abstract}
It has been suggested that the Newtonian gravitational force
may emerge as an entropic force from a holographic microscopic theory.
In this framework, the possibility is reconsidered
that Newton's gravitational coupling constant $G$
can be derived from the fundamental constants of
the underlying microscopic theory.
\vspace*{1\baselineskip}\newline
Journal-ref: Class. Quantum Grav. 28 (2011) 125003
\vspace*{.5\baselineskip}\newline
Preprint: IPMU10--0098,\; KA--TP--15--2010,\;arXiv:1006.2094 
\vspace*{.5\baselineskip}\newline
\end{abstract}



\maketitle

\section{Introduction}\label{sec:introduction}

Recently, Verlinde~\cite{Verlinde2010} has given a heuristic argument
of how space, inertia, and gravity could emerge from a microscopic theory in
a holographic approach~\cite{'tHooft1993,Susskind1994}.
Gravity would arise as a type of entropic force.
(Related ideas have been presented in, e.g.,
\cite{Jacobson1995,Padmanabhan2009}.)

Verlinde's discussion of Newton's law of gravity is particularly
elegant, as it directly gives an inverse-square law for the attractive
force between two macroscopic point masses $M_0$  and $M_1$.
Specifically, the force on the point mass
$M_0$ at position $\mathbf{X}_0$ due to an effective point mass
$\widetilde{M}_1$ at an effective position $\widetilde{\mathbf{X}}_1$
(the mass $\widetilde{M}_1$ corresponding to a spherical holographic screen)
is given by
\begin{eqnarray}\label{eq:F0grav}
\mathbf{F}_\textXXX{0,\,grav} &=& M_0\,\boldsymbol{A}_\textXXX{0,\,grav}
              =
G\,M_0\,\widetilde{M}_1\;
\big(\widetilde{\mathbf{X}}_1-\mathbf{X}_0\big)\big/
\big|\widetilde{\mathbf{X}}_1-\mathbf{X}_0\big|^3\,,
\end{eqnarray}
with $\boldsymbol{A}_\textXXX{0,\,grav}$ the acceleration of the mass $M_0$.

In this article, a previous suggestion~\cite{Klinkhamer2007}
is reconsidered that Newton's gravitational
constant $G$ can be derived from more fundamental constants of
nature, including a new fundamental length $l$
(see also \cite{Sakharov1967,Visser2002}
for a classic paper and a recent review).
The entropic explanation of Newtonian gravity then gives
a new interpretation of an earlier formula~\cite{Klinkhamer2007}
for the Newtonian
gravitational acceleration originating from a macroscopic point
mass. Moreover, having a new fundamental constant
$l$ may help in resolving a potential problem of Verlinde's approach
regarding the total entropy of a general equipotential screen.
Restricting the screen to a black-hole horizon,
this entropy can be used to perform
a model calculation of $G$ and to get a numerical estimate
for $l$ by connecting to the Bekenstein--Hawking
entropy~\cite{Bekenstein1973,Hawking1975}.
At the end of this article,
a few comments are presented on possible experiments
to determine this new fundamental constant $l$, if really existent.

\section{Nonfundamental $\boldsymbol{G}$}
\label{sec:Nonfundamental-G}

Consider the possibility that the true
fundamental constants of nature are $\hbar$, $c$, and $l^2$, where the last
constant has the dimensions of area. This suggests
(as mentioned in Sec.~2 of \cite{Klinkhamer2007}) that the
classical Newton constant $G$ arises from the appropriate \emph{ratio}
of the two quantum constants $l^2$ and $\hbar\,$:
\begin{equation}\label{eq:G}
G=f\,c^3\,l^2/\hbar\,,
\end{equation}
with a positive numerical factor $f \in\mathbb{R}^{+}$ to be
calculated from the microscopic theory. Note that, for $f=1$,
the fundamental length $l$ equals the
standard Planck length $l_{P}\equiv (\hbar\, G)^{1/2}\,c^{-3/2}$.

Expression \eref{eq:G} leads to the following structure of the
Newtonian gravitational
acceleration $A_\textXXX{grav} \equiv |\boldsymbol{A}_\textXXX{grav}|$ from a
point mass with a macroscopic value $M$ at a macroscopic distance $R\,$:
\begin{equation}\label{eq:Agrav}
A_\textXXX{grav}=G M/R^2 = c\;
\big( f\,Mc^2/\hbar\big)\, \big(l^2/R^2\big)\,,
\end{equation}
with all microscopic quantities indicated by lower-case symbols.
The structure on the right-hand side of \eref{eq:Agrav}
is suggestive: the fundamental velocity $c$
is multiplied by a mass-induced decay rate of space,
$f\,Mc^2/\hbar$ with coupling constant $f$,
and a geometric dilution factor, $l^2/R^2$.

\begin{figure}
\hfill
\includegraphics[width=0.875\textwidth]{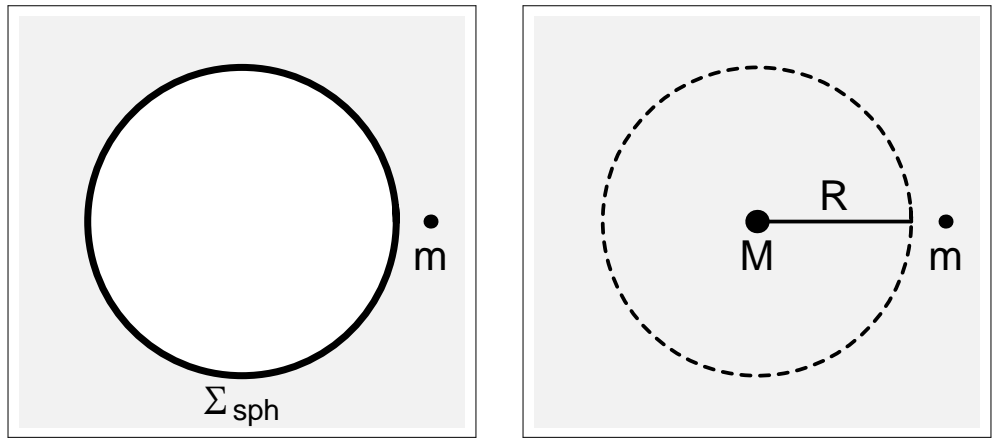}
\hspace*{-4mm}
\caption{\protect\underline{Left panel:}
Spherical holographic
screen $\Sigma_\textXXX{sph}$ with area $A=4\pi R^2$ and test mass $m$
in the emerged space (shaded) outside the screen~\cite{Verlinde2010}.
The screen $\Sigma_\textXXX{sph}$ has $N$ microscopic
degrees of freedom at an equilibrium temperature $T$ with
total equipartition energy $E=\textstyle{\frac{1}{2}}\, N\, k_{B}T$.
\protect\underline{Right panel:}
The gravitational effects of $\Sigma_\textXXX{sph}$
for the emergent space correspond, in leading order,
to those of a point mass $M=E/c^2$ located at the center
of a sphere with radius $R$. (The Schwarzschild radius
$R_\textXXX{Schw}\equiv 2 GM/c^2$ is considered to be negligible
compared to $R$ and cannot be shown in the right panel,
but the corresponding sphere would be a maximally-coarse-grained
screen with smallest possible area,
according to \cite{Verlinde2010}.)}
\label{fig:1}
\end{figure}

Precisely this structure can be seen to result from the reasoning
of Verlinde (see, in particular, Sec.~3 of \cite{Verlinde2010})
for a spherical holographic screen $\Sigma_\textXXX{sph}$
with area $A=4\pi R^2$ (Fig.~\ref{fig:1}):
\begin{eqnarray}\label{eq:Agrav-derivation}
A_\textXXX{grav}
&\stackrel{\textXXX{\textcircled{\sffamily\tiny 1}}}{=}&
2\pi\,   c\,  \big( k_{B}T /\hbar\big)
\nonumber\\[1mm]
&\stackrel{\textXXX{\textcircled{\sffamily\tiny 2}}}{=}&
4\pi\,  c \, \big(f\,\textstyle{\frac{1}{2}}\, N\, k_{B}T/\hbar\big) \, \big(f^{-1} / N\big)
\nonumber\\ [1mm]
&\stackrel{\textXXX{\textcircled{\sffamily\tiny 3}}}{=}&
4\pi\,  c \, \big(f\,E          /\hbar\big)\,  \big(l^2 / A\big)
\nonumber\\ [1mm]
&\stackrel{\textXXX{\textcircled{\sffamily\tiny 4}}}{=}&
c\,  \big(f\,M c^2      /\hbar\big) \, \big(l^2 / R^2\big)\,,
\end{eqnarray}
where step $\textcircled{\sffamily\scriptsize 1}$
relies on the Unruh effect~\cite{Unruh1976},
step $\textcircled{\sffamily\scriptsize 2}$ on trivial mathematics,
and step $\textcircled{\sffamily\scriptsize 3}$ on the
following relation between the effective
number $N$ of degrees of freedom of the holographic screen
and the area $A$ of the screen:
\begin{equation}\label{eq:N}
 N = f^{-1}\, A/l^2\,.
\end{equation}
Step $\textcircled{\sffamily\scriptsize 3}$ of \eref{eq:Agrav-derivation}
also assumes that the screen corresponds to a physical system in a state
of equilibrium (or close to it),
with a uniform distribution of the microscopic degrees of freedom
over the surface and equipartition
of the total energy $E$ over these degrees of freedom
(both properties being consistent with having a screen given
by a constant-curvature manifold, i.e., a spherical surface).
Somewhat surprisingly,  
Lorentz invariance is seen to play a
role in steps $\textcircled{\sffamily\scriptsize 1}$
and $\textcircled{\sffamily\scriptsize 4}$
of \eref{eq:Agrav-derivation}: implicitly as
the Unruh temperature ultimately traces back to the Lorentz invariance
of the Minkowski vacuum~\cite{Unruh1976} and explicitly through the
energy-mass equivalence $E\equiv M\,c^2$ from special relativity.

The several steps of \eref{eq:Agrav-derivation} constitute,
if confirmed by the definitive microscopic theory,
a \emph{derivation} of Newton's gravitational coupling
constant $G$ in the form \eref{eq:G}. The point of view of this article
is not to consider \eref{eq:Agrav-derivation} as mere dimensional
analysis but to take all numerical factors seriously.
In that spirit, there is the new insight
from \eref{eq:N} that, given the ``quantum of area'' $l^2$,
the inverse of the constant $f$ entering Newton's
constant \eref{eq:G} is related to the \emph{nature} of the
microscopic degrees of freedom on the holographic screen.
For example, an ``atom of space'' with ``spin'' $s_\textXXX{atom}$
gives $f^{-1}=2\,s_\textXXX{atom}+1$, but this
``spin'' need not be half-integer.
Still, the number of ``atoms'' needed to build-up the area $A$ is taken
to be an integer, given by the ratio of the area $A$ and the quantum $l^2$.

\section{Two types of entropy}
\label{sec:Two-types-of-entropy}

The introduction of \emph{two} quantum constants, $\hbar$ and $l^2$, may
also help to resolve a potential problem noted by Verlinde in Sec.~6.4 of
\cite{Verlinde2010}. There, he considers an equipotential screen
$\Sigma$ which is not a maximally-coarse-grained surface but is
nevertheless assumed to be in thermal equilibrium.
He, then, remarks that the required entropy $S_\Sigma$ appears to contradict
Bekenstein's upper bound~\cite{Bekenstein1981} on the entropy $S_{\,\Xi}$ of a
material system $\Xi$ with energy $E_{\,\Xi}$ and effective radius $R_{\,\Xi}$,
\begin{equation}\label{eq:S-Xi}
S_{\,\Xi}/k_{B} < 2\pi\;\hbar^{-1}\,E_{\,\Xi}\,R_{\,\Xi}/c\,.\
\end{equation}

With the new fundamental constant $l^2$, Verlinde's expression (6.41) for
$S_\Sigma$ is replaced by
\begin{equation}\label{eq:S-Sigma}
S_\Sigma/k_{B}
=\frac{1}{4}\,f^{-1}\;l^{-2}\,\int_\Sigma d A\,,
\end{equation}
which generalizes \eref{eq:N}.
Expressions \eref{eq:S-Xi} and
\eref{eq:S-Sigma} involve essentially different physics characterized by,
respectively, $\hbar$ and $l^2$ (see also the discussion of Sec.~2
in \cite{Klinkhamer2007} for
a generalized dimensionless action with $\hbar=0$ and $l^2>0$).
This observation would appear to support Verlinde's suggestion
that Bekenstein's bound may not apply to the holographic screen. Still, the
puzzle remains how these two types of entropy combine,
as they somehow must do in an appropriate limit.

\section{Model calculation of $\boldsymbol{G}$}
\label{sec:Model-calculation-of-G}

For a maximally-coarse-grained spherical surface (horizon) with area $A$,
the entropy \eref{eq:S-Sigma} reproduces the Bekenstein--Hawking
black-hole entropy~\cite{Bekenstein1973,Hawking1975}
\begin{equation}\label{eq:S-BH}
S_\textXXX{BH}/k_{B}= \frac{1}{4}\,A/(f\,l^2) = (1/4)\,N\,,
\end{equation}
where the number $N$ has already been defined by \eref{eq:N}.

Now, consider the ``atoms of space'' mentioned in the last paragraph
of Sec.~\ref{sec:Nonfundamental-G}.
The crucial new equation from \eref{eq:N} is then given by
\begin{equation}\label{eq:N-dof}
N=d_\textXXX{atom}\,N_\textXXX{atom}\,,
\end{equation}
with the physical interpretation of $l^2$ as the quantum of
area giving
\numparts\label{eq:N-atom-d-atom}
\begin{equation}\label{eq:N-atom}
N_\textXXX{atom}\equiv A/l^2
                \in \mathbb{N}_1 \equiv \{1,\, 2,\, 3,\, \ldots\}
\end{equation}
and the effective dimension of the internal Hilbert space
of a single ``atom of space'' taking values
\begin{equation}\label{eq:d-atom}
d_\textXXX{atom} \equiv f^{-1} \in \mathbb{R}^{+} 
\,.
\end{equation}
\endnumparts
The  physical picture, suggested by
the derivation \eref{eq:Agrav-derivation}, is
that the ``atoms of space'' have no translational degrees of freedom
but only internal degrees of freedom.

The number of configurations~\cite{'tHooft1993,Susskind1994} of these
distinguishable ``atoms of space'' is readily calculated:
\begin{equation}\label{eq:N-configurations}
N_\textXXX{config}
= \prod_{n=1}^{N_\textXXX{atom}}\,d_\textXXX{atom}
= \big(d_\textXXX{atom}\big)^{\,N_\textXXX{atom}}\,.
\end{equation}
Equating this number of configurations with the exponential of
the Bekenstein--Hawking entropy \eref{eq:S-BH}
while using \eref{eq:N-dof}
gives the following set of conditions:
\begin{equation}\label{eq:Natom-datom-condition}
\big(d_\textXXX{atom}\big)^{\,N_\textXXX{atom}}
= \exp\big[(1/4)\,d_\textXXX{atom}\,N_\textXXX{atom}\big]\,,
\end{equation}
for all positive integer values of $N_\textXXX{atom}$.
Remarkably, this infinite set of conditions reduces
to a single transcendental equation
for the effective dimension $d_\textXXX{atom}$,
\begin{equation}\label{eq:datom-condition}
\ln\,d_\textXXX{atom} = (1/4)\,d_\textXXX{atom}\,,
\end{equation}
which has two solutions:
\begin{equation}\label{eq:datom-solutions}
d_\textXXX{atom}^{\,(+)}\approx 8.613\, 169\, 456 \,, 
\quad
d_\textXXX{atom}^{\,(-)}\approx 1.429\, 611\, 825 \,, 
\end{equation}
where a $1$ ppb numerical precision suffices for the present
purpose. \footnote[3]{Condition \eref{eq:datom-condition}
would not be satisfied for any value of $d_\textXXX{atom}$
if the factor $1/4$ on the right-hand side,
which traces back to \eref{eq:S-BH}, were replaced
by an arbitrary number $g > 1/e$, with $e\approx 2.71828$
the base of the natural logarithm.
Note also that \eref{eq:datom-condition} rules out
$d_\textXXX{atom}=1$, corresponding to $f=1$ in
the original expression \eref{eq:G} for Newton's constant.}

Given $l^2$, there are then two possible values for
the gravitational coupling constant \eref{eq:G}:
\begin{equation}\label{eq:G-plusminus}
G_{\pm} = \big(d_\textXXX{atom}^{\,(\pm)}\big)^{-1}\; c^3\,l^2/\hbar\,.
\end{equation}
The detailed microscopic theory must tell which of the two values
from \eref{eq:datom-solutions} enters \eref{eq:G-plusminus}.
It could, for example, be that the microscopic theory
demands $d_\textXXX{atom} \geq 2$, selecting the larger
value $d_\textXXX{atom}^{\,(+)}$ in \eref{eq:datom-solutions}
and \eref{eq:G-plusminus}.

The experimental value $G_N$
of Newton's gravitational coupling constant is, of course, already
known \cite{Cavendish1798}, albeit with a rather large relative
uncertainty of $100$ ppm~\cite{MohrTaylorNewell2008}.
A more practical interpretation of result \eref{eq:datom-solutions}
is, therefore, to calculate two possible values for the
``quantum of area'':
\begin{eqnarray}\label{eq:l2-plusminus}
(l_{\pm})^2 &=&       d_\textXXX{atom}^{\,(\pm)}\; \big(l_{P}\big)^2
\approx
\left\{
\begin{array}{l}
  2.2498  \times 10^{-69}\;\textXXX{m}^2 \,,  
  \\
  3.7343 \times 10^{-70}\;\textXXX{m}^2  \,,  
\end{array}
\right.
\end{eqnarray}
with $l_{P}\equiv (\hbar\, G_N)^{1/2}/c^{3/2}
\approx  1.6162 \times 10^{-35}\;\textXXX{m}$
for $G_N= 6.6743(7)\;10^{-11}\;\textXXX{m}^{3}\;\textXXX{kg}^{-1}\;\textXXX{s}^{-2}$
~\cite{MohrTaylorNewell2008}.
The microscopic theory would, again, have to choose between
these alternative values.
For either choice, the implication would be that $l$ and $l_{P}$
are of the same order of magnitude.

Needless to say, the numerical estimates of \eref{eq:l2-plusminus}
are only indicative because of the extreme simplification of the
model calculation
(for example, merely ``tiles'' of a single size $l^2$ and
design $d_\textXXX{atom}$ have been used to cover the area $A$).
But, perhaps, the simplicity of the model is also its strength,
as long as the \emph{effective} quantum of area $l^2$
is considered and not the individual eigenvalues of the
area operator.

The real question is if this $l^2$ can be measured directly.
This question will be addressed in the next section.
Anticipating a positive outcome of that discussion
and looking far into the future,
note that the accurate measurement of one of the values of $l^2$
in \eref{eq:l2-plusminus} would allow for an equally accurate
calculation of $G$ from \eref{eq:G-plusminus}.
For example, measuring for $l^2$ the larger value
in \eref{eq:l2-plusminus} with a relative uncertainty of
$100$ ppb would give $G$ also with an uncertainty of
approximately $100$ ppb from \eref{eq:G-plusminus}
by use of the $d_\textXXX{atom}^{\,(+)}$ value
from \eref{eq:datom-solutions}, since $\hbar$ is
already known with an uncertainty of
$50$ ppb~\cite{MohrTaylorNewell2008}.

\section{Experiments}
\label{sec:Experiments}

As promised in the previous section,
let us briefly discuss the prospects of the experimental determination
of the factor $f$ in \eref{eq:G}, which may or may not be found to
agree with the inverse of one of the calculated values in
\eref{eq:datom-solutions}. Given the numerical values for $c$ and
$\hbar$ from nongravitational experiments, at least \emph{two}
gravity/spacetime measurements would be needed to disentangle $f$ and $l$.

The first measurement is, of course, provided by the Cavendish
experiment~\cite{Cavendish1798,MohrTaylorNewell2008}, which determines
the particular combination $f\,l^2$.

A second  measurement (without definite results, for the moment) can come
from cosmic-ray particle-propagation experiments probing Lorentz-violating
effects\footnote[4]{According to the discussion
in Sec.~\ref{sec:Nonfundamental-G}, it may be that the fundamental
theory is essentially Lorentz invariant. Still, there may be effects
from some type of spontaneous symmetry breaking of Lorentz invariance
(meaning that a particular ground-state solution breaks the symmetry),
which show up as modifications of the standard particle-propagation
properties.}
from a nontrivial small-scale structure of
spacetime~\cite{BernadotteKlinkhamer2006,KlinkhamerSchreck2008}. Such a
measurement may, in fact, determine $f^2\equiv (l_{P}/l)^4$, if
the average size of spacetime defects is set by $l_{P}$ and
their average  separation by $l$ (with $l > l_{P}$);
see the discussion of the paragraph
starting a few lines below Eq.~(10) in \cite{Klinkhamer2007}.
The value $f^2 \gtrsim 10^{-2}$ suggested by \eref{eq:datom-solutions}
would, however, be hard to reconcile with the
data (cf. \cite{BernadotteKlinkhamer2006,KlinkhamerSchreck2008}
and references therein).

A third type of measurement (entirely in the domain of
\emph{Gedankenexperiments}) could try to isolate pure-quantum-gravity effects
of (primordial) gravitational waves. Such a measurement would
only depend on $l^2$, if
the generalized dimensionless action of \cite{Klinkhamer2007}
is relevant.

A fourth type of measurement (also in the domain of
\emph{Gedankenexperiments}) would look for quantum modifications
of Newton's gravitational acceleration \eref{eq:Agrav} by a
multiplicative factor $\big[1-\widetilde{a}\,l^2/R^2\big]$,
where the dimensionless number $\widetilde{a}$ would trace  
back to a logarithmic correction of the entropy \eref{eq:S-Sigma},
as pointed out in \cite{ModestoRandono2010}.
More generally, an entropy modification
$S(A)= (1/4)\,k_{B}\,l_{P}^{\, -2}\,\big[A+l^2\,\widetilde{s}(A/l^2)\big]$,
for some dimensionless function $\widetilde{s}$ of $A/l^2$, 
would give a correction factor $\big[1+l^2\,d\widetilde{s}/dA\big]$
for Newton's gravitational force.
A measurement of such a modification of the force
could, in principle, be used
to determine $l^2$, if the function $\widetilde{s}(A/l^2)$
is nontrivial and known from theory.

Each of the last three possible experiments relies on a crucial assumption
(indicated by occurrence of the word `if')
and is, therefore, not yet conclusive in determining the value of $l^2$.

\section{Conclusion}
\label{sec:Conclusion}

The two most interesting results of this article are the following.
The first is that the interpretation  
of the Newtonian acceleration \eref{eq:Agrav}
as a mass-induced decay rate of space
(together with a geometric dilution factor) may be explained by
a Verlinde-type derivation (\ref{eq:Agrav-derivation}) relying on
the Unruh temperature and holography.
The second is the single transcendental equation \eref{eq:datom-condition},
which allows for an explicit calculation of the numerical factor
$f \equiv \big(d_\textXXX{atom}\big)^{-1}$ entering
expression \eref{eq:G} for Newton's gravitational constant,
where the microscopic theory is still needed to choose
between the two possible values \eref{eq:datom-solutions}.

Having a calculated value for $f$ in the $G$ formula \eref{eq:G}
is, of course, only of interest if $l^2$ can be determined directly
(the numerical values of $G$, $\hbar$, and $c$ are already known).
The experiments discussed in Sec.~\ref{sec:Experiments} are
suggestive but, for the moment, still inconclusive, because each
experiment involves one or more assumptions.
The main outstanding task, therefore, is to design an
experiment, real or imaginary, which allows for an unambiguous
determination of the quantum-gravity length scale $l$,
independent of the value of the Planck length $l_{P}$
[even though, in the end,
both may turn out to have approximately the same numerical value,
as suggested by the calculated numbers \eref{eq:l2-plusminus}].

The first version of the present article 
was released on June 10, 2010. Since then,
it has been shown~\cite{Sahlmann2010} that a more
sophisticated tiling than the one used in
Sec.~\ref{sec:Model-calculation-of-G} can produce a
single  transcendental equation which gives a \emph{unique}
physical value for $l^2$. The numerical values for $l^2$
from two such tiling models are both
approximately equal to $2.6  \times 10^{-69}\;\textXXX{m}^2$,
which is only $20\,\%$ above the maximal value found here.
More importantly, the quantity $l^2$ of these
models~\cite{Sahlmann2010} would
correspond to the true minimal quantum of area.

\hfill
\section*{Acknowledgements}

The author thanks D. Easson for pointing out \cite{Verlinde2010} and
S. Hellerman, H. Sahlmann, L. Smolin, W. Unruh, and E. Verlinde
for discussions. He also gratefully acknowledges
the hospitality of the IPMU during the month of May 2010.
This work was supported in part by the
World Premier International Research Center Initiative
\mbox{(WPI Initiative),} MEXT, Japan.

\newpage

\end{document}